\documentclass[floatfix,twocolumn]{revtex4-1}
\usepackage{graphicx}
\usepackage{color}
\usepackage{natbib}
\usepackage{hyperref}
\usepackage{cleveref}
\usepackage{morefloats}

\begin{document}
\title{Formation and stability of a hollow electron beam in the presence of a plasma wake field driven by an ultra-short electron bunch}
\author{F. Tanjia$^1$, R. Fedele$^1$, S. De Nicola$^{2,1}$, T. Akhter$^1$, D. Jovanovi\'{c}$^3$}
\affiliation{{\small $^1$ Dipartimento di Fisica, Universi\`{a} di Napoli ``Federico II" and INFN, Napoli, Italy\\
$^2$ CNR-SPIN, Complesso Universitario  di Monte S'Angelo, Napoli, Italy\\ $^3$ Institute of Physics, University of Belgrade, Belgrade, Serbia}}

\begin{abstract}
A numerical investigation on the spatiotemporal evolution of an electron beam, externally injected in a plasma in the presence of a plasma wake field, is carried out. The latter is driven by an ultra-short relativistic axially-symmetric femtosecond electron bunch. We first derive a novel Poisson-like equation for the wake potential where the driving term is the ultra-short bunch density, taking suitably into account the interplay between the sharpness and high energy of the bunch. Then, we show that a channel is formed longitudinally, through the externally injected beam while experiencing the effects of the bunch-driven plasma wake field, within the context of thermal wave model. The formation of the channel seems to be a final stage of the 3D evolution of the beam. This involves the appearance of small filaments and bubbles around the longitudinal axis. The bubbles coalesce forming a relatively stable axially-symmetric hollow beam structure.
\end{abstract}

\maketitle

\section{Introduction}
In this manuscript, we are going to investigate the physical conditions to generate the hollow structure by means of a plasma wake field (PWF) excitation mechanism similar to the laser wake field (LWF) excitation. We use a relativistic high energy ultra-short electron bunch as a driver, whose time duration ranges from sub-picoseconds to femtoseconds, and a moderately long charged particle beam as a driven system, whose time duration ranges from ($10^3 - 10^2$) femtoseconds. The hollow structure results from the interaction of the PWF generated by the ultra-short bunch with the driven beam. Here, we study numerically the evolution of the driven beam within the framework of the Thermal Wave Model (TWM) for charged particle beam propagation \cite{Fedele1991, Fedele1992a, Fedele2012a, Fedele2011, Tanjia2011, Fedele2014}, where a Schr\"{o}dinger-like equation governs the longitudinal spatiotemporal evolution of a complex wave function, whose squared modulus is proportional to the beam density. The adopted model equations constitute a pair of coupled partial differential equations comprising a Poisson-like equation and the Schr\"odinger-like equation, constructed in the following way. 

We first consider a cylindrically symmetric relativistic ultra-short bunch moving along the $z-$axis at the velocity $\beta c \hat{z}\, (\beta\simeq 1)$. We denote with $\rho_b(z,r,t)$ the number density of the bunch where $r$ is the cylindrical radial coordinate and $t$ is the time coordinate. In order to get an equation for the wake potential energy, we perform the coordinate transformation $\xi =z-\beta ct, \, r' = r, \, \tau=ct$. Under this transformation the linearized Lorentz-Maxwell fluid equations of the ``bunch+system" can be reduced to the following Poisson-like equation
{\small\begin{equation}
\left(\frac{\partial^2}{\partial\xi^2}+1\right) \left(\frac{1}{\gamma_0^2}\frac{\partial^2}
{\partial\xi^2}+\nabla_\perp^2-1\right)U_w=-\left(\frac{1}{\gamma_0^2}\frac{\partial^2}{\partial\xi^2}-1\right)\frac{\rho_b}{n_0\gamma_0},\label{3h}\
\end{equation}}
where $U_w(r,\xi)=-q\,\Omega(r,\xi)/m_0\gamma_0c^2$ is the dimensionless wake potential energy, $\Omega=\left(\beta A_{1z}-\phi_1\right)$ is the dimensional wake potential, $A_{1z}$ is the longitudinal component of the perturbation of vector potential $\mathbf{A}_1$, $\phi_1$ is the perturbations of scaler potential $\phi$, respectively, $\gamma_0$ is the leading order term of the relativistic factor $\gamma=(1-\beta^2)^{-1/2}$, $k_{pe}\equiv\omega_{pe}/c$, $\omega_{pe}=(4\pi n_0e^2/m_0)^{1/2}$ is the electron plasma frequency, and $q=-e$ is the charge of the bunch. To obtain Eq. (\ref{3h}) we first observed that $\nabla_\perp = \partial/\partial r=\partial/\partial r'$ and further assumed that, on the fast time scale, $\partial/\partial\tau=0,$ which imposes the quasi-electrostatic approximation. Therefore, Eq. (\ref{3h}) relates $U_w(\xi,r)$ to $\rho_b(\xi,r)$ during the early times  (note that we have, for simplicity, replaced $r'$ by $r$).
\begin{figure}
  \centering
  \includegraphics[width=7cm]{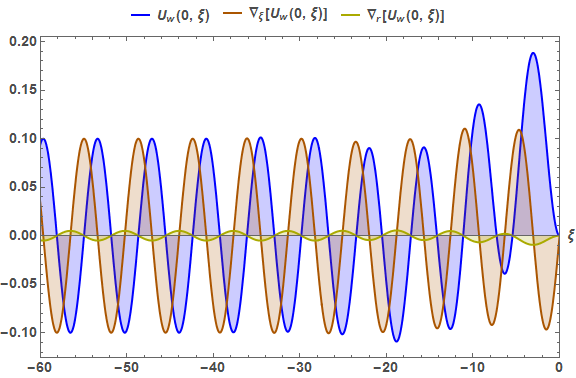}\\
  \caption{Longitudinal evolution of the normalized wake potential energy $U_w$ (blue curve) and the corresponding longitudinal and radial gradients $\nabla_{\xi}U_w$ (orange curve) and $\nabla_{r}U_w$ (green curve) respectively, in the vicinity of the longitudinal axis ($r\rightarrow 0$) for dimensionless bunch length $\sigma_z\rightarrow k_{pe}\sigma_z\simeq 10^{-2} \,(\simeq\,0.1\,\mu m)$ and spot size $\sigma_\perp\rightarrow k_{pe}\sigma_\perp\simeq 5.5 \,(\simeq 50\,\mu m)$.}\label{fig1}
\end{figure}
We assume that a second cylindrically symmetric beam (i.e., \textit{witness} or \textit{driven} charged-particle beam), is launched toward the plasma wake along $z-$axis to experience the effects of the PWF produced by the driving bunch (i.e., ultra-short bunch). Therefore, the longitudinal spatiotemporal evolution of the driven beam  manifests on longer time scales and in quantum-like domain of TWM is provided by the following Schr\"{o}dinger-like equation, where the quasi-electrostatic assumption is removed, viz., 
{\small\begin{equation}
i\epsilon\frac{\partial\psi}{\partial\tau}=-\frac{\epsilon^2}{2}\frac{\partial^2\psi}{\partial\xi^2}+U_w\psi,\label{4a}\
\end{equation}}
\noindent where $\psi(r,\xi,\tau)$ is the complex wave function called \textit{beam wave function} and $\epsilon$ is the thermal beam emittance. Note that, we have made all the variables dimensionless with respect to $k_{pe}$, viz., $\tau\rightarrow k_{pe}\tau$, $\xi\rightarrow k_{pe}\xi$, $r\rightarrow k_{pe}r$, $\epsilon\rightarrow\epsilon k_{pe}$, and $\psi=\psi/k_{pe}^{3/2}$. Note also that $\rho_b'(r,\xi,\tau)=N|\psi(r,\xi,\tau)|^2$, where $N$ is the total number of driven beam particles.

Equation (\ref{3h}) differs from the standard one of PWF theory \cite{Chen1985} and contains second and fourth order derivatives with respect to $\xi$. To obtain this equation , we have taken into account carefully the longitudinal sharpness of the bunch compared to its high energy conditions i.e., value of $\gamma_0$. Note that, our assumption of ultrashort electron bunch leads to the condition that the bunch length is much less than the plasma wavelength, i.e., $k_{pe}\sigma_z\ll 1$, where $\sigma_z$ is the bunch length. Therefore, we are looking forward to study a regime where the ultra sharpness of the bunch length ($\partial/\partial\xi$) compensates the smallness of $1/\gamma_0$ in such a way that the term $\left|1/\gamma_0 \left(\partial/\partial\xi\right)\right|$ in Eq. (\ref{3h}) could not be neglected and be comparable to $1$. This leads to the condition $1/\gamma_0\approx k_{pe}\sigma_z \ll 1$. To study the behavior of the wake potential from Eq. (\ref{3h}), this condition must be satisfied for any set of parameters.

The pair of Poisson-like and Schr\"odinger-like equation, i.e., Eqs. (\ref{3h}) and (\ref{4a}), respectively, resembles the Zakharov-like coupled system of equations. It describes the spatiotemporal evolution of the driven beam while interacting with the plasma wake that has been generated by the ultra-short driving bunch. In the next section, we will present the numerical results of this coupled system of equations.
\begin{figure}
  \centering
 \includegraphics[width=8cm]{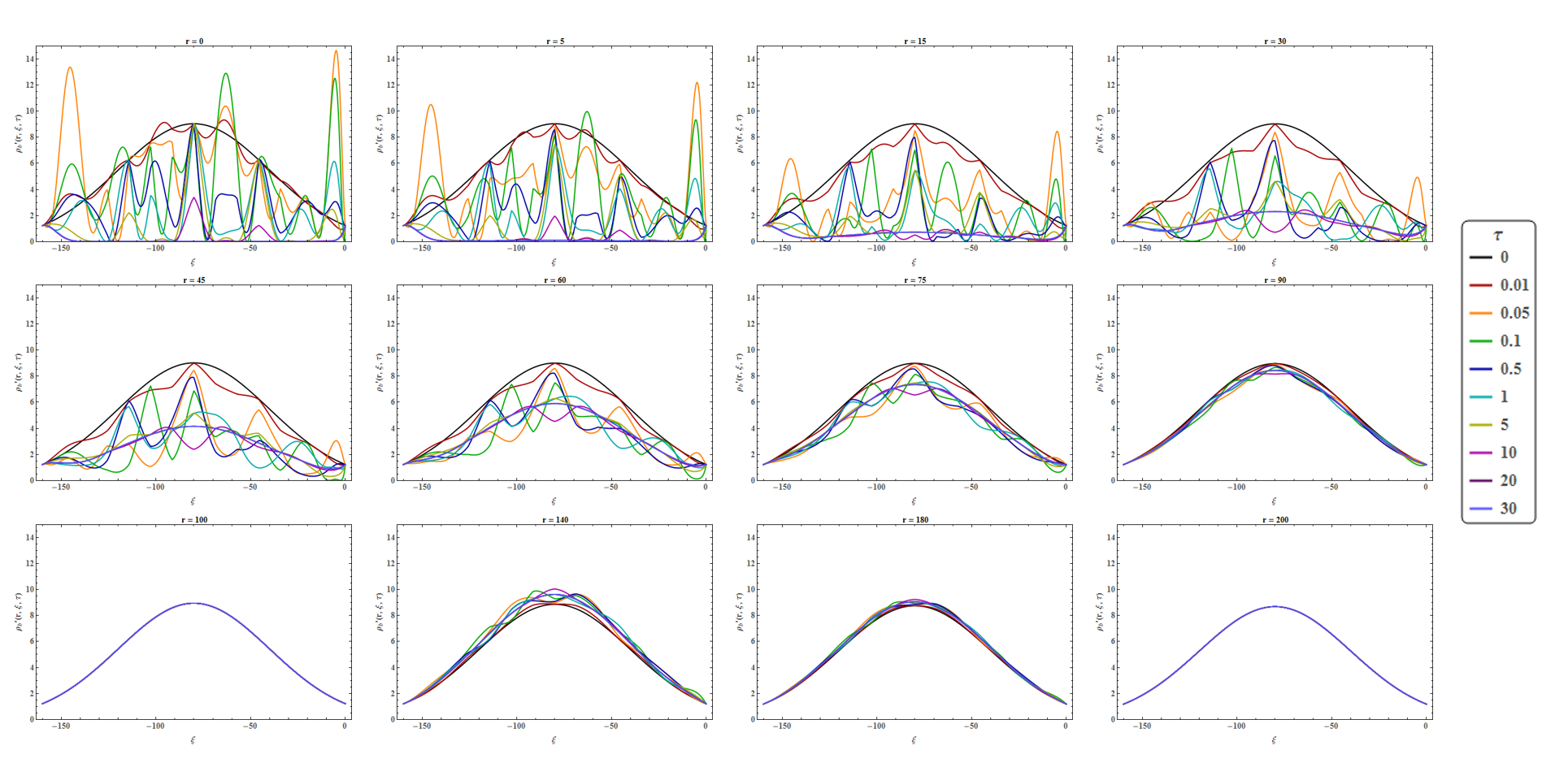}\
  \caption{Longitudinal oscillation of $\rho_b'(r,\xi,\tau)$ at different $r$ for $\sigma_z'=40$ and $\sigma_\perp'=1000$ with a longitudinal offset $\bar{\xi}=80$ and $\epsilon=10^{-3}$.}\label{faa1}
\end{figure}
\begin{figure}
  \centering
 \includegraphics[width=8cm]{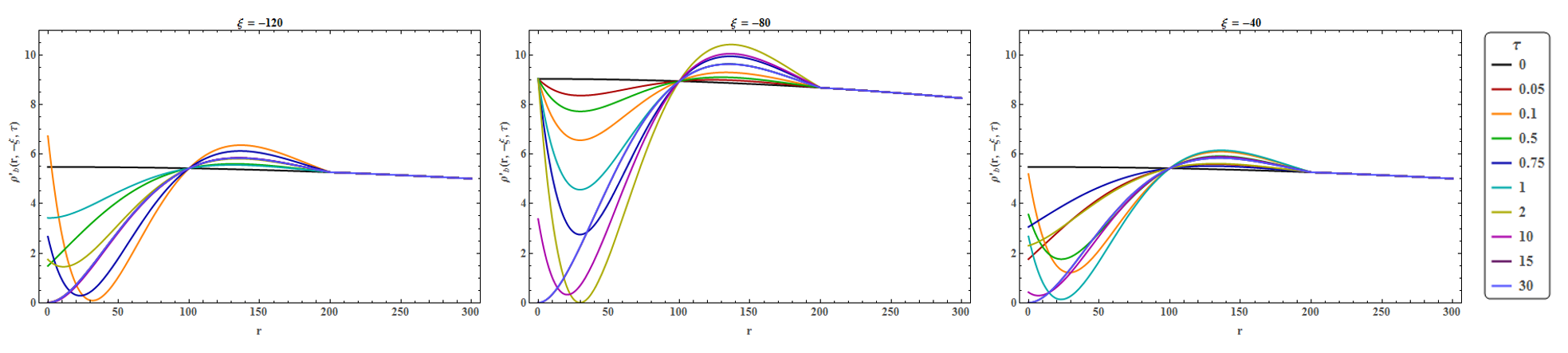}\
  \caption{Radial oscillation of $\rho_b'(r,\xi,\tau)$ at different $\xi$ for $\sigma_z'=40$ and $\sigma_\perp'=100$ with a longitudinal offset $\bar{\xi}=80$ and $\epsilon=10^{-3}$.}\label{faa2}
\end{figure}
\section{Numerical Results}
As we have already pointed out, Eq. (\ref{3h}) that governs the spatiotemporal evolution of the wake potential has been numerically integrated by assuming the Gaussian distribution in cylindrical symmetry, viz., $\rho_b(r,\xi)=n_b \exp\left[-\left(\frac{\xi^2} {2k_{pe}^2\sigma_z^2} + \frac{r^2}{k_{pe}^2\sigma_\perp^2}\right)\right]\,,$ where $\sigma_z$ and $\sigma_\perp$ are the bunch length and spot size of the driving bunch, respectively. The spatial distribution of the dimensionless wake potential energy $U_w(r,\xi)$ [ normalized by $\left(n_b/n_0\gamma\right)$] is plotted as a function of  dimensionless $\xi$ and $r$.

Figure \ref{fig1} shows the longitudinal evolution of the normalized wake potential energy $U_w$ and the corresponding longitudinal and radial gradients $\nabla_{\xi}U_w$ and $\nabla_{r}U_w$ respectively, in the vicinity of the longitudinal axis ($r\rightarrow 0$). The corresponding values of the bunch length and spot size (both normalized by $k_{pe}$) are considered as $\sigma_z\rightarrow k_{pe}\sigma_z\simeq 10^{-2} \,(\simeq\,0.1\,\mu m)$ and $\sigma_\perp\rightarrow k_{pe}\sigma_\perp\simeq 5.5 \,(\simeq 50\,\mu m)$, respectively. The normalization factor for $U_w$ is $n_b/n_0\gamma\simeq 10^{-7}$ ($\gamma_0\simeq 10^{3}$, $n_0\simeq 10^{17}$ cm$^{-3}$, and  $n_b\simeq 10^{14}$ cm$^{-3}$). All the parameters have been chosen in such a way to satisfy the condition $1/\gamma_0\approx k_{pe}\sigma_z \ll 1$. The wake potential energy $U_w (\xi,0)$ (blue line) and the corresponding longitudinal and radial gradients, i.e.,  $\nabla_\xi [U_w(\xi,0)]$ (orange line) and $\nabla_r [U_w(\xi,0)]$ (green line), respectively, exhibit regular oscillations  along $\xi$. The longitudinal gradient is much greater than the radial one.
\begin{figure}
  \centering
  \includegraphics[width=6cm]{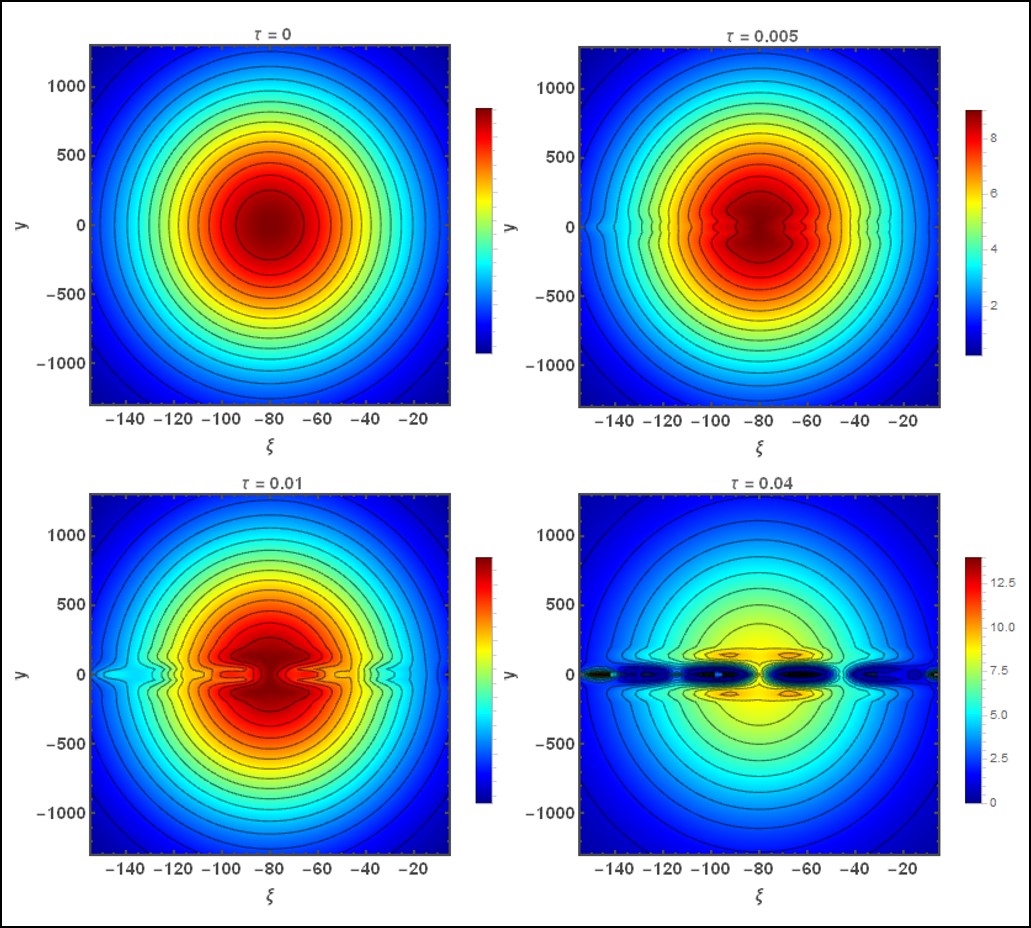}\\
  \vspace{0.24cm}
  \includegraphics[width=6cm]{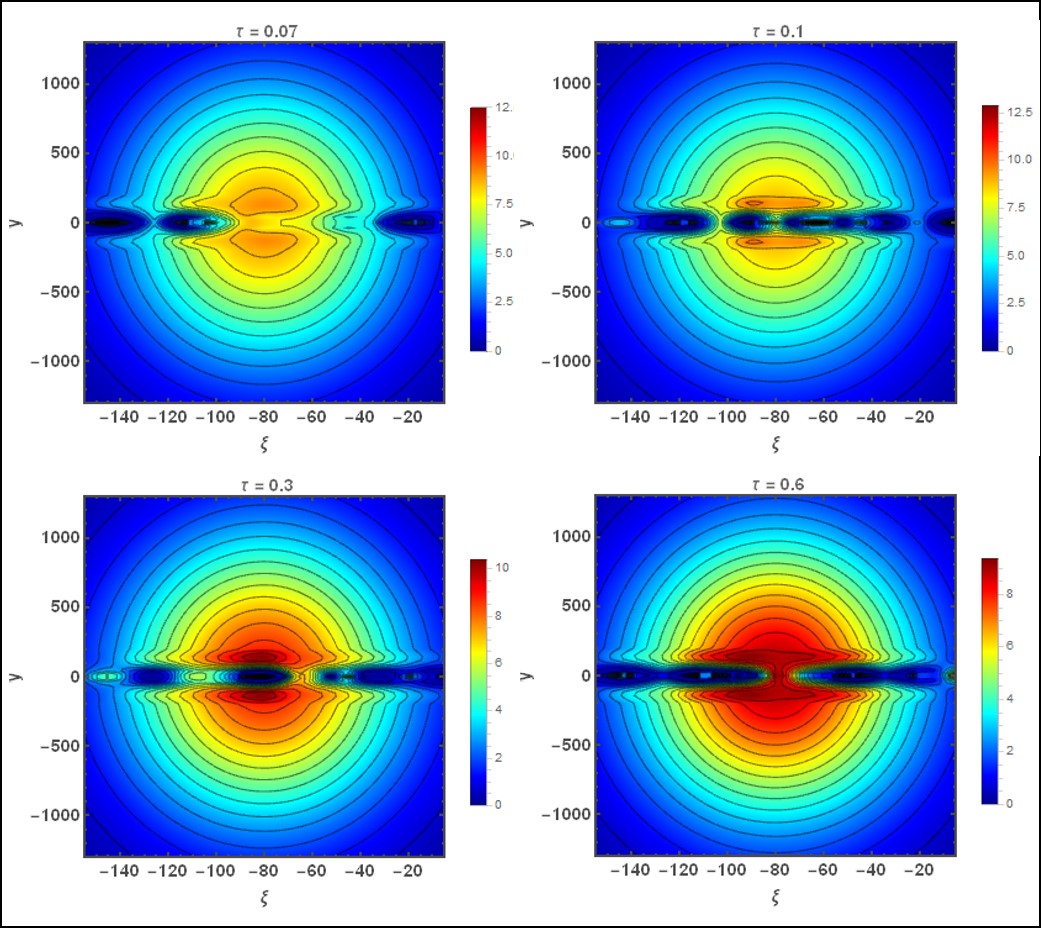}\
  \caption{Evolution of $\rho_b'(x,y,\xi,\tau)$ at the depth of $x=0$ for $\sigma_z'=40$ and $\sigma_\perp'=1000$ with a longitudinal offset $\bar{\xi}=80$ and $\epsilon=10^{-3}$ from $\tau=0-0.04$ (up) and $\tau=0.07-0.6$ (down). The scale of the plot legend has been normalized by the factor $10^{13}$.}\label{fa2-1}
\end{figure}
Next, for the driven beam, we chose an initial normalized dimensionless Gaussian profile similar to the driving bunch but longitudinally off-set by the normalized length $\bar{\xi}\rightarrow k_{pe}\bar{\xi}$ at $\tau=0$, of the form $\psi(r,\xi,0)=\frac{1}{\sqrt{\sqrt{2\pi}\sigma_z'\pi \sigma_\perp'^2}}\, \exp\left[-\left(\frac{(\xi+\bar{\xi})^2} {4\sigma_z'^2} + \frac{r^2}{2\sigma_\perp'^2}\right)\right]\,,$ where $\sigma_z'$ and $\sigma_\perp'$ are the beam length and spot size, respectively, that are normalized by $k_{pe}$, viz., $\sigma_z'\rightarrow k_{pe}\sigma_z'$ and $\sigma_\perp'\rightarrow k_{pe}\sigma_\perp'$. For this initial beam profile, we numerically solve the Schr\"odinger-like equation (\ref{4a}), in which $U_w$ is the output of numerical solution of Eq. (\ref{3h}). For the driven beam, we chose $\sigma_z'=40$ and $\sigma_\perp'=1000$ with an offset $\bar{\xi}=80$ and $\epsilon=10^{-3}$. Note that, we have chosen the normalized dimensionless beam length $\sigma_z'$ in such a way that it is comparable to the wake field wavelength ($\sim 100\, \mu$m). In this conditions, we can assume that the self-interaction is negligible. Therefore, it is justified that in Eq. (\ref{4a}), we did not take into account the interaction of the driven beam on itself (self-interaction). In the next sections, we analyze the spatiotemporal evolution of the driven beam density $\rho_b'(r,\xi,\tau)=N |\psi(r,\xi,\tau)|^2$ in different dimensions, i.e., 1D, 2D, and 3D, respectively. 
\begin{figure}[htb!]
  \centering
  \includegraphics[width=6cm]{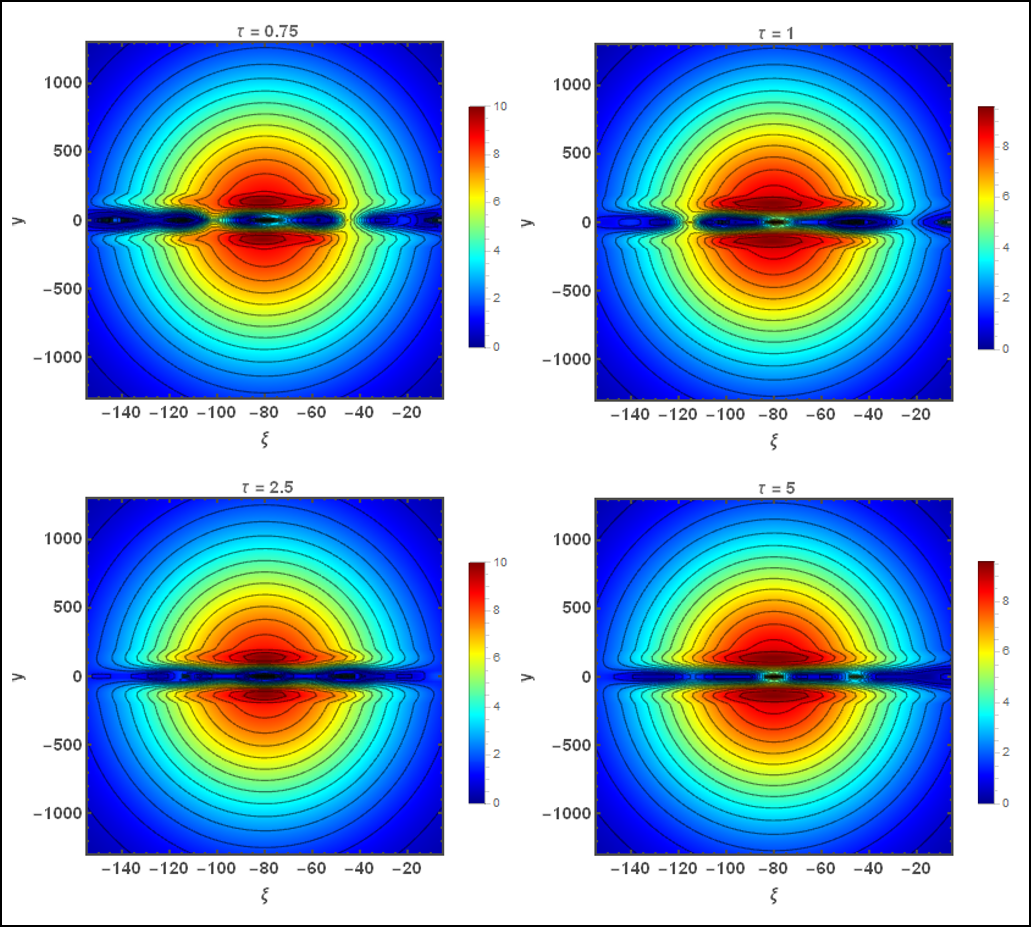}\\
  \vspace{0.24cm}
  \includegraphics[width=6cm]{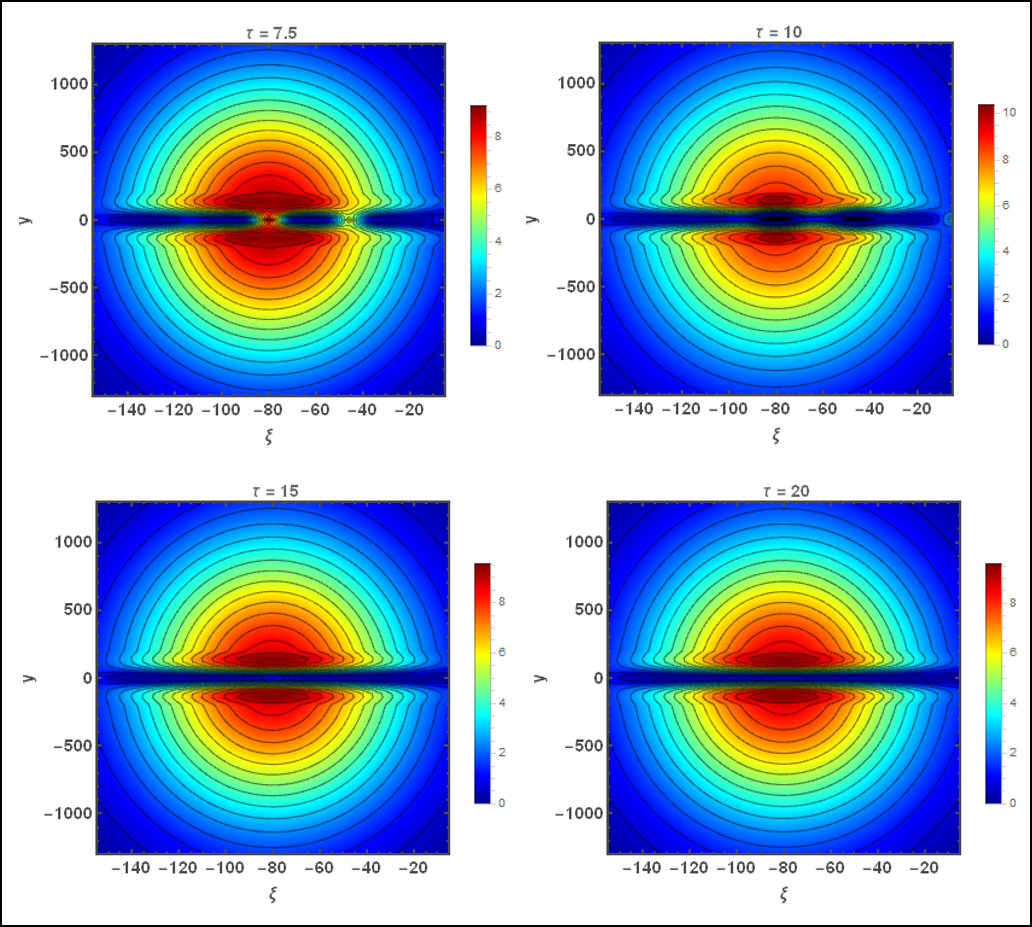}\
  \caption{Evolution of $\rho_b'(x,y,\xi,\tau)$ at the depth of $x=0$ for $\sigma_z'=40$ and $\sigma_\perp'=1000$ with a longitudinal offset $\bar{\xi}=80$ and $\epsilon=10^{-3}$ from $\tau=0.75-5$ (up) and $\tau=7.5-20$ (down). The scale of the plot legend has been normalized by the factor $10^{13}$.}\label{fa2-2}
\end{figure}
\subsection{Density oscillations in 1D}
Figures \ref{faa1} and \ref{faa2} show the longitudinal oscillations of $\rho_b'(r, \xi, \tau)$ as a function of $\xi$ at given $r$ and $\tau$, and radial oscillations as a function of $r$ at given $\xi$ and $\tau$, respectively, for $\sigma_z'=40, \sigma_\perp'=1000$ with an initial Gaussian profile. The longitudinal oscillations are vivid at the radial origin ($r\approx 0$) and start to decrease as $r$  increases, as shown in  Figure \ref{faa1}. Note that, with increasing $tau$, starting from $r=0$ till $100$ we observe decrements of the total particles through oscillations whilst between $r=100$ and $r=200$ we observe increment. The profiles at $r=100$ and $r=200$ for any $\tau$ overlap to reconstitute the initial condition. For a fixed $r$, the longitudinal density oscillations with respect to $\xi$ are rapid in early $\tau$ and then reduce as $\tau$ increases. The radial oscillations of $\rho_b'(r, \xi, \tau)$ for fixed $\xi$ and $\tau$ are clearly visible in Figure \ref{faa2}. We have chosen the values of $\xi$ in such a way that: the first ($\xi=-120$) is located in one $\sigma_z'$ left from the Gaussian pick; the second ($\xi=-80$) is located at the Gaussian pick; finally, the third ($\xi=-40$) is located in one $\sigma_z'$ right from the Gaussian pick. We see that, starting from a radial Gaussian profile, which is flatter than the longitudinal one, this profile starts to oscillate with increasing $\tau$ making a dip-hump structure. This dip-hump structure exhibits the existence of two nodes, namely,  two fixed points which are not affected by the oscillations for all $\tau$. For the three different cases of $\xi$, they seem to be located in the same value of $r$ (about $r= 100$ and $r=200$) for all $\tau$, which we also observe in the longitudinal oscillations through the reconstruction of initial condition. The first node is located, with respect to $r =0$, at the distance corresponding to the aperture of the dip. The second node is located, with respect to the first node, at the distance corresponding to the effective physical extent of the hump. For all the chosen values of $\xi$, we observe that the density oscillations of the dip-hump structure stabilise  after some time about $\tau=20$. Starting approximately from this time, all the curves seem to be overlapping each other. However, the density oscillations that we observe in this 1D case is not the complete scenario. We further analyse the 2D case in the next section.
\subsection{Density oscillations in 2D}
Expressing the radial coordinate in terms of  the Cartesian coordinates $x$ and $y$ as $r=\sqrt{x^2+y^2}$,  we follow the spatiotemporal evolution of the beam density $\rho_b'(x,y,\xi,\tau)=N |\psi(x,y,\xi,\tau)|^2$ at different depths of $x$. We divide the time-like variable $\tau$ in different segments depending on the nature of the evolution. The first segment is the very early times $\tau=0-0.6$, when the system evolves very quickly. Next is the time frame of $\tau=0.75-20$  in which system evolves slower compared to the early times. We continue to follow the evolution till a final time $\tau=200$, when system evolves very slowly towards stabilisation. We have observed several interesting phenomena while following the evolution, whose descriptions are given below.
\begin{figure}[htb!]
  \centering
  \includegraphics[width=6cm]{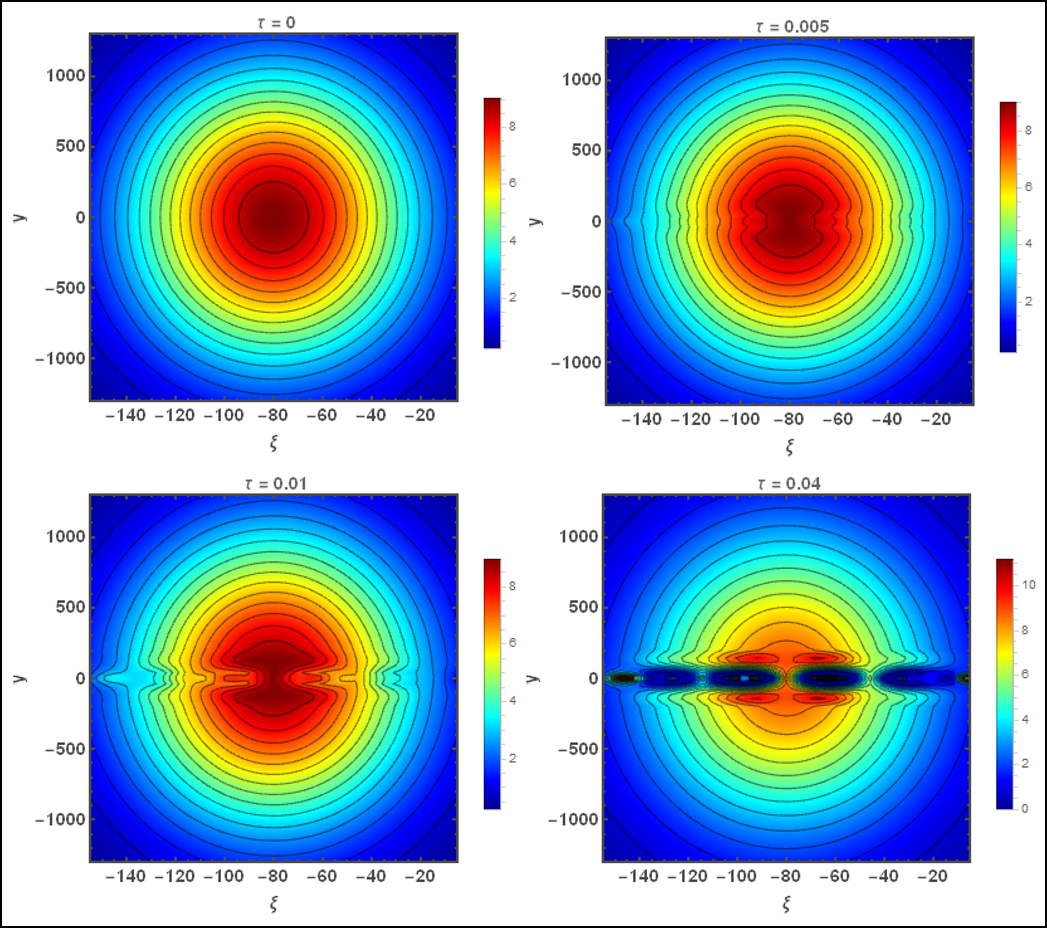}\\
  \vspace{0.24cm}
  \includegraphics[width=6cm]{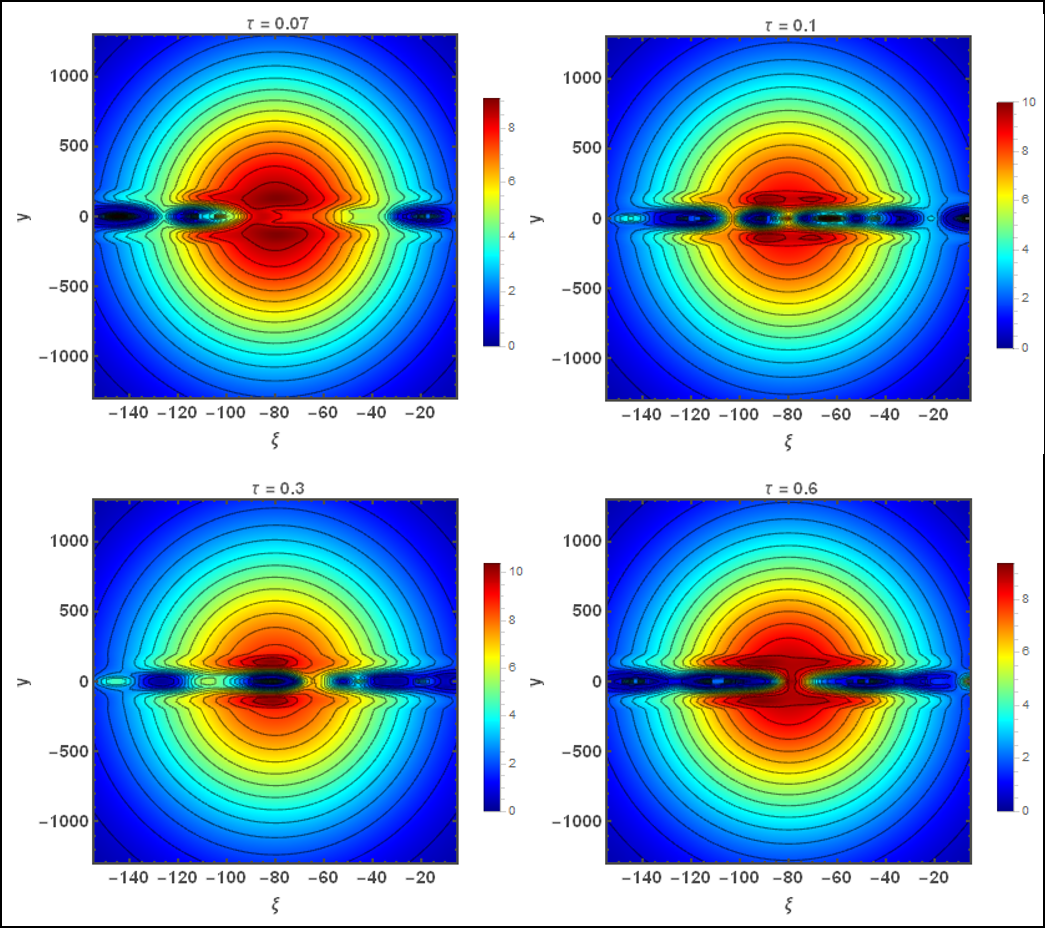}\
  \caption{Evolution of $\rho_b'(x,y,\xi,\tau)$ at the depth of $x=5$ for $\sigma_z'=40$ and $\sigma_\perp'=100$ with a longitudinal offset $\bar{\xi}=80$ and $\epsilon=10^{-3}$ from $\tau=0-0.6$ (up) and $\tau=0.08-0.5$ (down). The scale of the plot legend has been normalized by the factor $10^{13}$.}\label{fa3-1}
\end{figure}
\subsubsection{Formation of filaments and voids}
These features occur in the very early stage of times $\tau= 0-0.6$, in which the change in the initial profile starts to happen very quickly for different depths of $x$. Here we showed two different planes namely, $x=0$ and $x=5$. Figures \ref{fa2-1} and \ref{fa3-1} show the evolution of $\rho_b'(x,y,\xi,\tau)$ for $\sigma_z'=40$ and $\sigma_\perp'=1000$ with a longitudinal offset $\bar{\xi}=80$ and $\epsilon=10^{-3}$, at the depth of $x=0$ and $x=5$, respectively, during the early stage period ($\tau= 0-0.5$) of evolution, dividing it into two panels i.e., $\tau= 0-0.04$ (up) $\tau= 0.07-0.5$ (down). We observe a deformation of the core of the initial profile. The pattern in the core located around the longitudinal axis $y\approx 0$ evolves experiencing a sort of longitudinal elongation. The outer part of the core for larger values of $y$ evolves experiencing a contraction along $\xi$. The core around ($y\approx 0$) evolves assuming cigar shaped structures in both longitudinal edges that progressively leave the original core and appear located inside the voids. Then, while forming the cigar structures, the rest of the core deforms into a bone-like structure oriented vertically (along $y$) extending between two symmetric part of the edges. The waist of this structure lies along the longitudinal axis at $y=0$. In summary, the density plot of the original core evolves showing to be composed of two different types of structure, i.e., cigar shaped /filamentary and bone-like structures till the evolution reaches the time around $\tau=0.04$. Within this time, these two structures have different density of particles. After this time (around $\tau=0.07$), the above structures start to deform in such a way to appear longitudinally asymmetric. At $\tau= 0.6$, it is already evident that the bone-like structure starts to break the waist and the density of the particles in the filaments reduces. The evolution of the pattern in both the planes (corresponding to Figures \ref{fa2-1} and \ref{fa3-1}) are almost identical, however in $x=5$ plane the distribution of particles are different than in $x=0$ plane. The main difference that we observer is that, in the latter the sharp cigar shaped filaments are not as visible as the previous one. For the structures at the same $\tau$ but located in the above two different planes, the densities are complementary. This means that an increasing of density in one plane corresponds to a decreasing density in the other plane for a fixed $\tau$. This is in agreement with 1D observations that we have done in the previous section showing the existence of both the longitudinal and radial density oscillation.
\subsubsection{Coalescence of voids and channelling}
\begin{figure}[htb!]
  \centering
  \includegraphics[width=6cm]{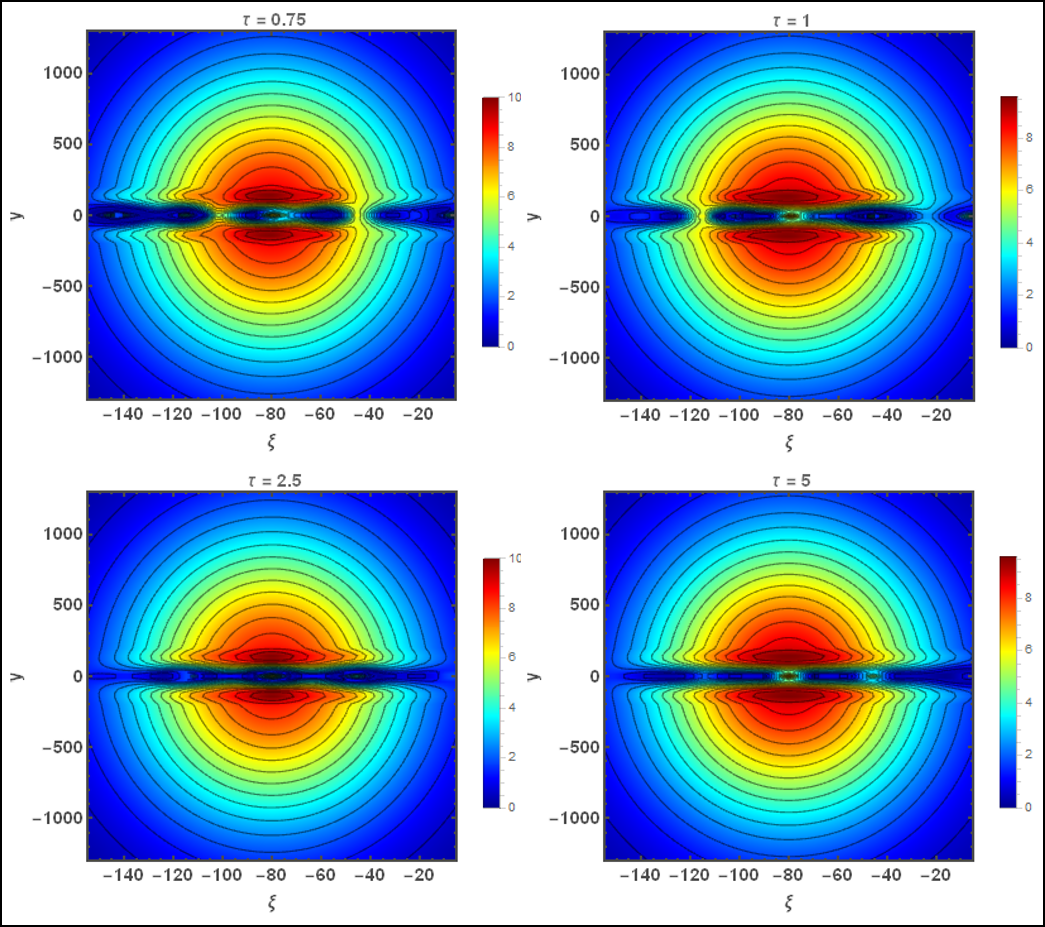}\\
  \vspace{0.24cm}
  \includegraphics[width=6cm]{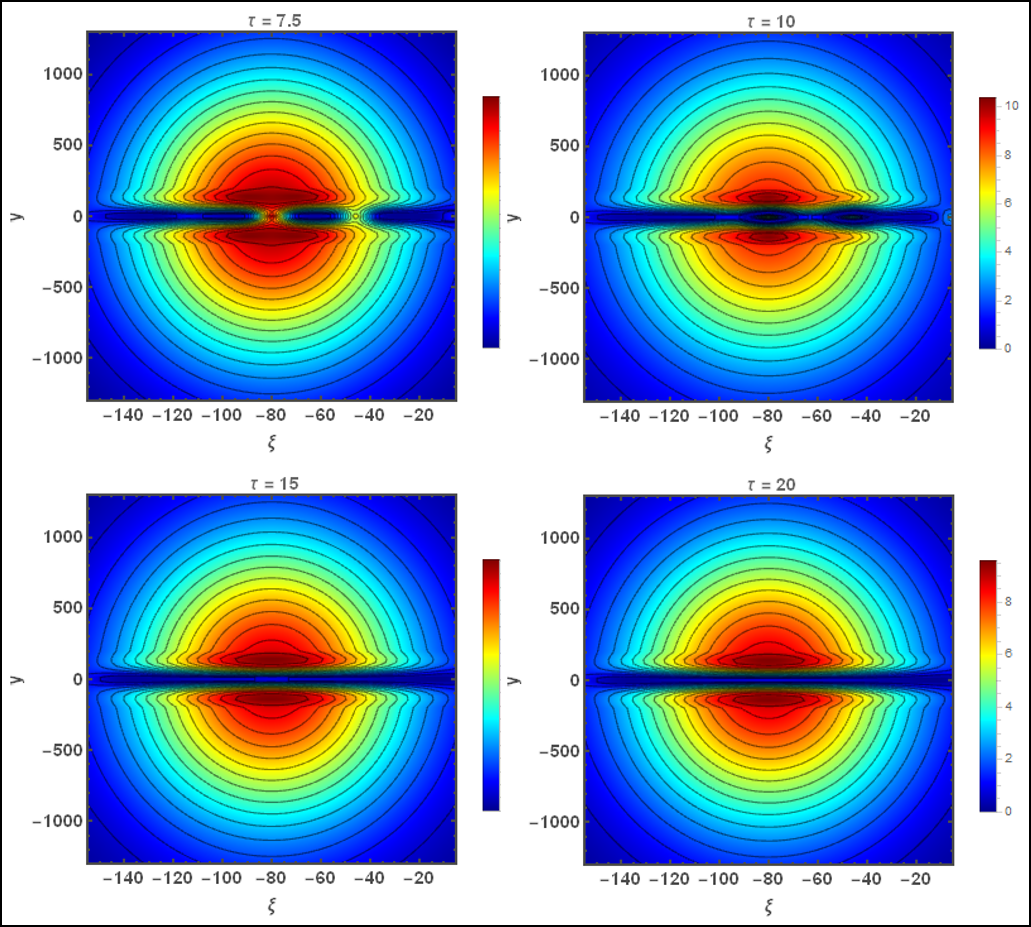}\
  \caption{Evolution of $\rho_b'(x,y,\xi,\tau)$ with the presence of both $U_w$ and $U_s$ at the depth of $x=5$ for $\sigma_z'=40$ and $\sigma_\perp'=100$ with a longitudinal offset $\bar{\xi}=80$ and $\epsilon=10^{-3}$ from $\tau=0.75-5$ (up) and $\tau=7.5-20$ (down). The scale of the plot legend has been normalized by the factor $10^{13}$.}\label{fa3-2}
\end{figure}

We continue to follow the evolution of $\rho_b'(x, y, \xi, \tau)$ for the same set of parameters in further time segments, i.e., $\tau= 0.75-20$ again dividing it into two panels [$\tau= 0.75-0.5$ (up) $\tau= 7.5-20$ (down)]. The density evolves qualitatively in a similar way for both the planes $x=0$ and $x=5$, shown in Figures \ref{fa2-2} and \ref{fa3-2}, respectively. In both the planes, the evolution continues in such a way that the region around the longitudinal axis becomes more empty (about $\tau=1$). Long and short voids are still visible containing filaments inside, as we observed in the previous stages ($\tau=0-0.6$, see Figures \ref{fa2-1} and \ref{fa3-1}) as well. The core progressively breaks into two quasi-symmetric disjoint regions while the voids coalesce. The coalescence continues until the time $\tau=15$ forming a channel in the core.
\subsubsection{Stable hollow beam formation: 3D structures}
\begin{figure}
  \centering
  \includegraphics[width=7.5cm]{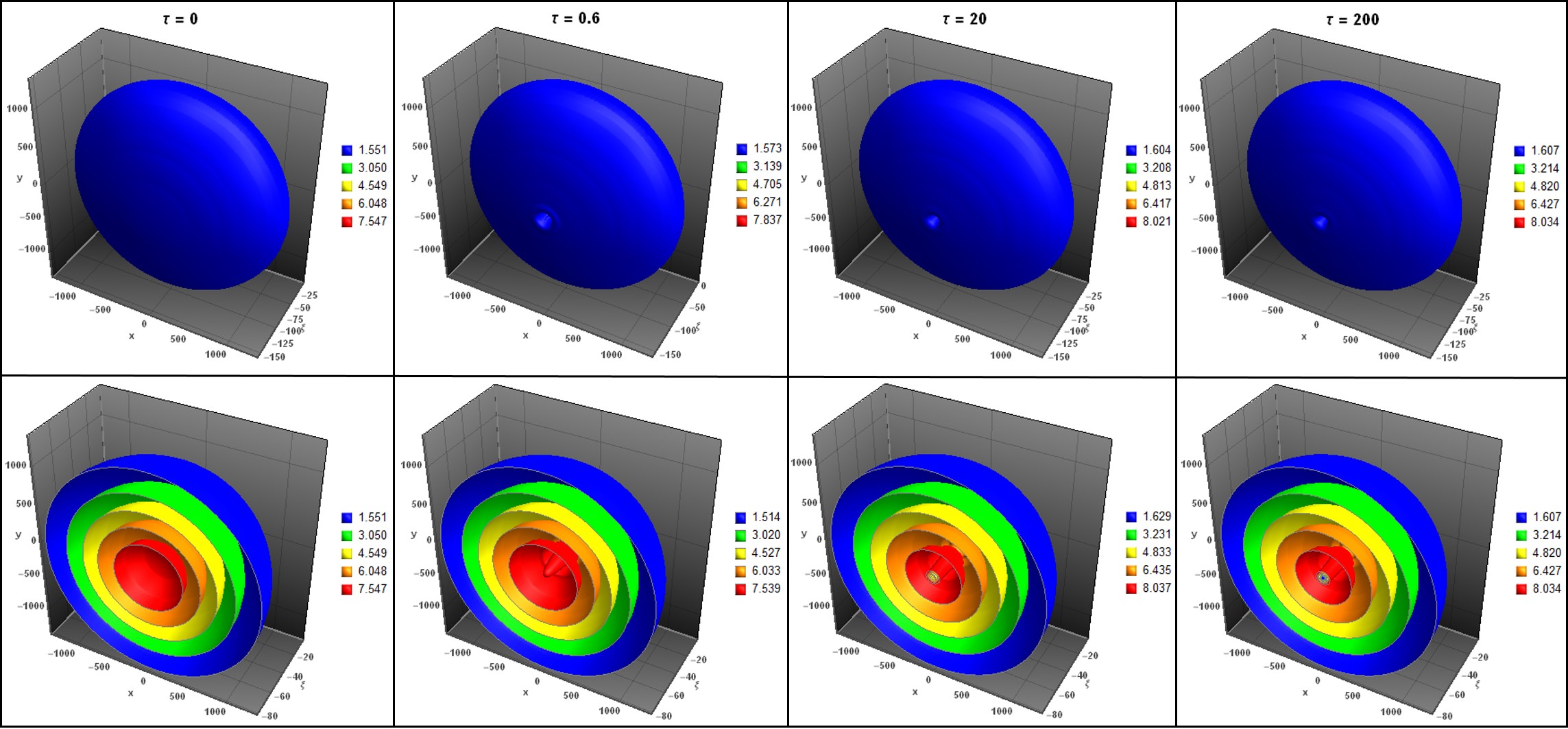}\
  \caption{3D representation of  $\rho_b'(x,y,\xi,\tau)$ for $\sigma_z'=40$ and $\sigma_\perp'=100$ with a longitudinal offset $\bar{\xi}=80$ and $\epsilon=10^{-3}$.}\label{fa4}
\end{figure}

\begin{figure}
  \centering
  \includegraphics[width=7.5cm]{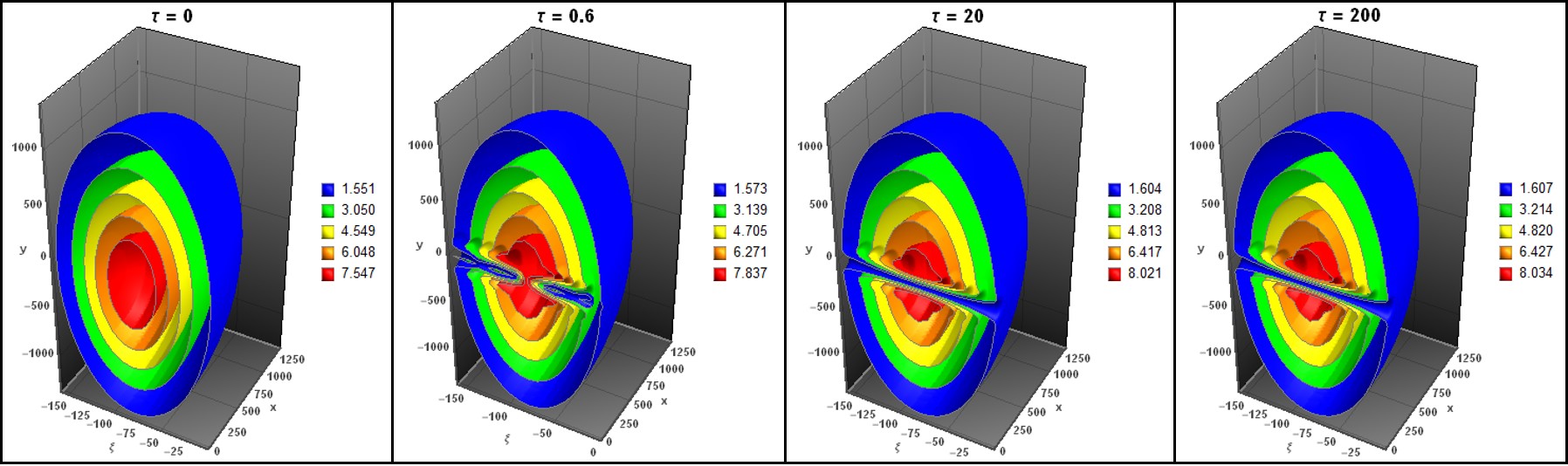}\
  \caption{3D contour representation of  $\rho_b'(x,y,\xi,\tau)$ for $\sigma_z'=40$ and $\sigma_\perp'=100$ with a longitudinal offset $\bar{\xi}=80$ and $\epsilon=10^{-3}$.}\label{fa5}
\end{figure}

We continue to follow the evolution till a final time $\tau=200$. To complete the scenario, instead of 2D density representation, we present here the 3D contour representation of $\rho_b'(x,y,\xi,\tau)$ for the same set of parameters provided in previous sections. The gradual formation of the hollow structure  are represented in the 3D contour representation of $\rho_b'(x,y,\xi,\tau)$ in Figures \ref{fa4} and \ref{fa5}. Figure \ref{fa4} represents the evolution of the beam with different $\tau$ (for $\tau=0,0.6,20,200$, respectively) in two different perspectives. The first row shows the beam in the total space of $\xi$, $x$, and $y$ initially which is a Gaussian. With the increasing $\tau$ the profile becomes the hollow structured in the core. The figure of the second row of Figure \ref{fa4} shows a cross sectional view cut in $\xi$ from which the inside structure of the hollow beam is visible. Instead, Figure \ref{fa5} shows another perspective of the beam cutting a cross section along $x$. This shows clearly the formation of hollow structure. In the beginning at $\tau=0$ the beam is perfectly Gaussian. As time increases, around $\tau=0.6$ it starts to have small bubbles at $y=0$ along $\xi$. The outer blue layer corresponds the lower values of beam density. However, all the particles are not completely extinguished from the core, as could be seen that along with the mostly blue layer, the presence of few dots of yellow and red layers corresponding to higher values of beam density. In the third and fourth column of  Figure \ref{fa5}, we see two equally divided structures hollowed in the middle at $y=0$. This structure completes the perspective seen in the single cross sectional density representations at $x=0$ and $x=5$ in Figures \ref{fa2-1} $-$ \ref{fa3-2}. In summary, starting from around $\tau=20$ the structure becomes completely hollow as can be seen clearly from Figure \ref{fa5}.
\section{Conclussions and remarks}\label{conclusions}
We have described the interaction between the PWF generated by a relativistic, femtosecond ultra-short (compared to the plasma wavelength) electron bunch and a moderately long (compared to the plasma wavelength) electron beam. The bunch has been assumed to be traveling in an unmagnetized cold plasma in overdense regime. This has been accomplished by carrying out a numerical investigation of the evolution of the driven beam within the framework of TWM. By using Lorentz-Maxwell system of equations for ``plasma+beam" system, in the quasi-static approximation, we have obtained a novel 3D Poisson-like equation, relating the plasma wake potential to the bunch density, that accounts for the arbitrary sharpness of the bunch. This equation is a fourth-order partial differential equation with respect to the longitudinal variable. By assuming a cylindrically symmetric bi-Gaussian 3D bunch density profile, we have solved numerically the Poisson-like equation for the wake potential, which in turn provides the potential term of the Schr\"odinger-like equation for the driven beam. The latter governs longer time scale spatiotemporal evolution of the driven beam wave function and therefore the beam density. The further numerical analysis of the Schr\"odinger-like equation has lead to a very rich collection of results concerning the interaction between wake field and the driven beam, such as void and filament formation, coalescence of voids. They have been characterized by means of 1D, 2D, and 3D representations. It has been observed that these effects originate and evolve in a very short time and contribute on a longer time scale to the hollow beam formation and its asymptotic stability. Remarkably, all these effects are density manifestations of the beam density oscillations in the longitudinal as well as radial directions. These oscillations are coupled and obey the conservation of the number of beam particles and, during the intermediate stages, involve the \textit{bucket formation} as a collective manifestation of the motion of each single particle of the driven beam in presence of the plasma wake potential well. 





%
%


\begin{thebibliography}{7}
\expandafter\ifx\csname natexlab\endcsname\relax\def\natexlab#1{#1}\fi
\expandafter\ifx\csname url\endcsname\relax
  \def\url#1{\texttt{#1}}\fi
\expandafter\ifx\csname urlprefix\endcsname\relax\def\urlprefix{URL }\fi

\bibitem[{Chen et~al.(1985)Chen, Dawson, Huff, and Katsouleas}]{Chen1985}
Chen, P., Dawson, J.~M., Huff, R.~W., Katsouleas, T., 1985. Acceleration of
  electrons by the interaction of a bunched electron beam with a plasma. Phys.
  Rev. Lett. 54, 693--696.

\bibitem[{Fedele and Miele(1991)}]{Fedele1991}
Fedele, R., Miele, G., 1991. A thermal wave model for relativistic charged
  particle beam propagation. Nuovo Cim. D 13, 1527--1544.

\bibitem[{Fedele and Shukla(1992)}]{Fedele1992a}
Fedele, R., Shukla, P.~K., 1992. Self-consistent interaction between the plasma
  wake field and the driving relativistic electron beam. Phys. Rev. A 45,
  4045--4049.

\bibitem[{Fedele et~al.(2012)Fedele, Tanjia, {De Nicola}, Jovanovi\'c, and
  Shukla}]{Fedele2012a}
Fedele, R., Tanjia, F., {De Nicola}, S., Jovanovi\'c, D., Shukla, P.~K., 2012.
  Quantum ring solitons and nonlocal effects in plasma wake field excitations.
  Phys. Plasmas 19, 102106, and references therein.

\bibitem[{Fedele et~al.(2011)Fedele, Tanjia, {De Nicola}, Shukla, and
  Jovanovi\'c}]{Fedele2011}
Fedele, R., Tanjia, F., {De Nicola}, S., Shukla, P.~K., Jovanovi\'c, D., 2011.
  Self consistent thermal wave model description of the transverse dynamics for
  relativistic charged particle beams in magnetoactive plasmas. In: Becoulet,
  A., Hoang, T., Stroth, U. (Eds.), Proc. 38th EPS Conf. Plasma Phys. Vol. 35G.
  Eur. Phys. Soc., Strasbourg, France.

\bibitem[{Fedele et~al.(2014)Fedele, Tanjia, Jovanovi\'c, {De Nicola}, and
  Ronsivalle}]{Fedele2014}
Fedele, R., Tanjia, F., Jovanovi\'c, D., {De Nicola}, S., Ronsivalle, C., 2014.
  Wave theories of non-laminar charged particle beams:from quantum to thermal
  regime. J. Plasma Phys. 80, 133--145, and references therein.

\bibitem[{Tanjia et~al.(2011)Tanjia, {De Nicola}, R.Fedele, Shukla, and
  Jovanovi\'c}]{Tanjia2011}
Tanjia, F., {De Nicola}, S., R.Fedele, Shukla, P.~K., Jovanovi\'c, D., 2011.
  Quantumlike description of the nonlinear and collective effects on
  relativistic electron beams in strongly magnetized plasmas. In: Becoulet, A.,
  Hoang, T., Stroth, U. (Eds.), Proc. 38th EPS Conf. Plasma Phys. Vol. 35G.
  Eur. Phys. Soc., Strasbourg, France.

\end{thebibliography}

\end{document}